# Parametric Simulation of Dynamic Glazing System with Changeable NIR Response

**Laura Hinkle[1], Neda Ghaeili Ardabili[1], Julian Wang[1,2*]**

[1] Department of Architectural Engineering, Pennsylvania State University, University Park, PA, USA 16802

[2] Materials Research Institute, Pennsylvania State University, University Park, PA, USA 16802

\* Corresponding Author: Julian Wang, julian.wang@psu.edu

**Abstract**. Windows are one of the main contributors to building energy consumption and emerging dynamic window technologies offer improved performance. Specifically, NIR-focused window technologies are desirable in climates that consume both heating and cooling energy. However, the whole building energy effects of changeable NIR response of building windows have not been captured, largely due to the lack of an appropriate energy simulation method and NIR-focused window modeling. This study focuses on developing a simulation method that enables the comprehensive evaluation of the whole building energy effects of dynamic NIR modulations. Using an *EnergyPlus* EMS-based parametric framework, annual energy savings were estimated for a switchable between-glass built-in system across three representative cities in ASHRAE climate zones 3, 4, and 5. This NIR-focused technology yielded energy savings of up to 19%. The results demonstrate the effects of NIR-focused window technologies on heating and cooling loads in different climates.

## 1. Introduction

Energy-efficient building windows are crucial to achieving low-energy buildings. The current methods for developing building glazing systems with high performance come from two primary directions: either structural or spectral design. In theory, from the structural design perspective, adding layers of windowpane separated by insulating barriers will allow the "minimum" level of conductive heat transfer. One cannot simply add an unlimited number of thermal barriers and windowpane layers, as this would result in unacceptable transparency, haze, and other practical issues related to thickness, heaviness, installation, maintenance, etc. Consequently, research has focused on addressing spectral characteristics in terms of thermal radiation and solar energy utilization. For temperatures of interest to building energy efficiency, thermal radiation flux occurs in four basic bands: ultraviolet irradiance, visible light, infrared irradiance from the sun, and longwave irradiance from the interior or outdoor temperatures. About 50% of the power of solar irradiance is infrared, so regulating a window's spectral properties to control the passage of thermal radiation in these ranges can significantly improve building energy efficiency. The most widely-accepted solution today is with low-emissivity (low-e) coatings, which may modulate the solar heat gains, especially the infrared portion, and then enhance the overall thermal performance of windows while retaining high visible transmittance (VT). The design of low-e coatings could be illustrated based on the "ideal window" concept (**Fig. 1a**) in terms of two types of idealized spectral curves – Type A and Type B for cooling- and heating-dominated climates, respectively. As is well known, the visible solar spectrum is considered essential to building occupants (e.g., for their well-being, health, work efficiency, etc.) and useful for other forms of energy such as electric light. Both types of low-e coatings would therefore permit high visible light transmission ($T_{vis}$) in the visible spectrum region. Furthermore, it would control non-visible radiative emissions for minimum energy consumption and reasonable indoor comfort. Type A is characterized by minimum infrared transmission (near and long-wave regions). Type B allows maximum near-infrared penetration while maintaining low transmittance of the long-wave thermal radiation [1].

Solar heat gain coefficient (SHGC) is normally used to guide the selection of glazing systems in different weathers. High-solar-gain (high SHGC) low-e coatings and low-solar-gain (low SHGC) low-e coatings have been designed and particularly used for buildings located in cold and hot climates, respectively. Figs 1b and 1c present two representative low-e coatings in the current market for two types of idealized curves. The key difference between them lies in the solar near-infrared (NIR) transmittance, which is about 50% for the high SHGC low-e coating and 5% for the low SHGC low-e coating. However, in most climates in the U.S., because both heating and cooling are necessary, dynamically switchable solar NIR transmittance is highly desirable for building energy savings.

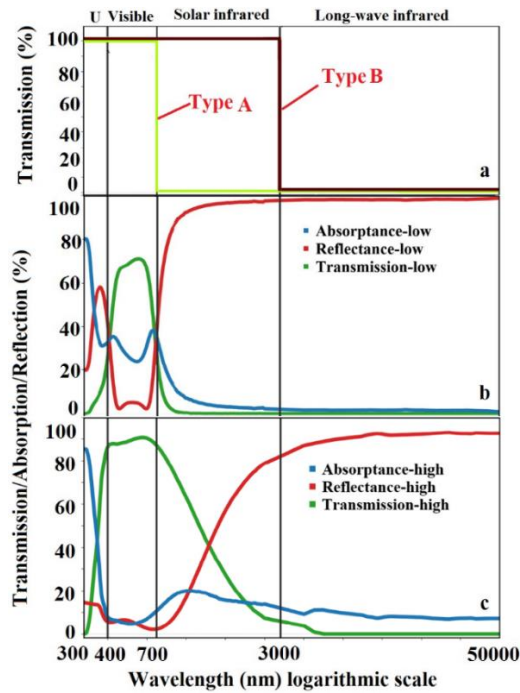

**Figure 1.** (a) Idealized coatings: Type A and Type B is characterized by transmittance in specific wavebands and ideally for summer and winter, respectively. (b) Spectra of representative low-SHGC low-e coatings. (c) Spectra of representative high-SHGC low-e coatings.

Despite significant efforts in pursuing such smart windows, challenges remain, attributed to uncoupled solar heat control in the visible and NIR wavelengths, complexity in installation and construction, and ongoing maintenance and operation costs. Recent developments in smart materials may pave a new way to enable the dynamic switch of solar NIR response in different seasons. In particular, some thermal responsive materials and structures can be tuned to have critical transition temperatures within the ambient temperature ranges, and then these materials can be integrated into dynamic glazing systems to mechanically switch the presence of the high SHGC low-e coating and the low SHGC low-e coating. There are several emerging techniques towards such dynamics, and while it is evident that some dynamic structures may positively influence building window energy performance, we could not identify any studies that investigated the comprehensive annual building energy effect of solar NIR-focused dynamic technologies. In this study, therefore, we explore the method to simulate and evaluate the window's dynamic NIR modulation effects on the heating and cooling loads of buildings.

## 2. Literature Review

Several emerging technologies towards changeable NIR response include kirigami-based built-in blind systems [2], coupled wood bilayer shading systems [3], and thermal bimetal window systems [4]. Li et al [5] described applications of shape-memory polymers, proposing shape-memory polymer-driven panel bending systems to reduce building lighting energy. Zhou et al conducted monthly energy simulations to demonstrate the performance of a 3D grating shape structure with W-doped VO2 polymer to control NIR transmittance [6]. Similarly, Ke et al [7] designed a smart window based on plasmonic-enhanced reconfigurable elastomer in order to modulate solar transmittance. Despite these advances toward dynamic NIR response, the whole building dynamic NIR modulation effects on heating and cooling loads of buildings have not been captured due to the lack of an appropriate simulation method.

Parametric energy simulation has proved a powerful tool in simulating emerging window technologies [8], [9]. Many previous works have leveraged parametric energy simulation to study thermochromic glazing [10]–[12] and other smart window technologies [13]–[15], as well as kinetic shading devices [16]–[18]. When evaluating current methods to dynamically adjust solar heat gain coefficient (SHGC) and visible transmittance (VT), most solutions are complex, requiring low-voltage equipment or actuators. More importantly, due to the lack of comprehensive simulation analysis, the potential energy savings by the NIR modulation of glazing systems are yet to be known.

## 3. Methodology

In this work, an ultra-thin, transparent *kirigami* sheet (~0.1mm thick) is proposed and sandwiched in the air gap between double panes (**Fig. 2**). By operating its opening area ratios in response to outdoor air temperature, the window can passively switch between the high- and low-solar-gain low-e coating states. *Kirigami,* through its cuts and reconnections, brings an extra, previously unattainable level of design, dynamics, and deployability to open and close thin sheets even at the architectural scale. Here, the low-solar-gain low-e coating is applied on the outer surface of the central *kirigami* sheet, while the outer surface of the inner window pane is coated with the high-solar-gain low-e coating (see **Fig. 2a**). The *kirigami* sheet can be bent, folded, and recovered reversibly with high durability. With a pattern of discrete cuts, the free faces of the *kirigami* sheet can undergo out-of-plane folding upon the forces of shrinkable tapes to form openings (**Fig. 2b**). As such, the area ratio of the openings (=opening area/sheet area) of the overall *kirigami* sheet can be automatically modulated from 0 to 0.8 in summer and winter, respectively. The working mechanism is illustrated in **Fig. 2c**. *In summer*, because of the tape shrinkage under a high temperature, the closed, transparent *kirigami* sheet has a uniform and complete coverage and in turn, a strong solar near-infrared reflection. This will enable a low SHGC of overall building windows, thus saving cooling energy use. *In winter*, the low temperature in the air gap drives the expansion of the tapes, which subsequently makes the sheet's segmental faces undergo out-of-plane folding. Up to 80% of openings on the sheet will be formed and enable high solar NIR transmittance. The high SHGC value offsets the building's heating energy use. This study seeks to quantify the annual building energy savings from implementing this mechanism in ASHRAE climate zones 3, 4, and 5. As illustrated above, with the proposed switchable between-glass built-in system, it is possible to achieve dynamic solar gain designs with one product. In order to quantify the potential energy savings, this system was simulated using parametric energy simulations in EnergyPlus across multiple climates. Notably, this work is not to explore the mechanism or materialization of the actuation of dynamic low-e coating behaviors but rather focuses on developing a simulation method enabling the comprehensive evaluation of the whole building energy effects of the dynamic NIR modulations.

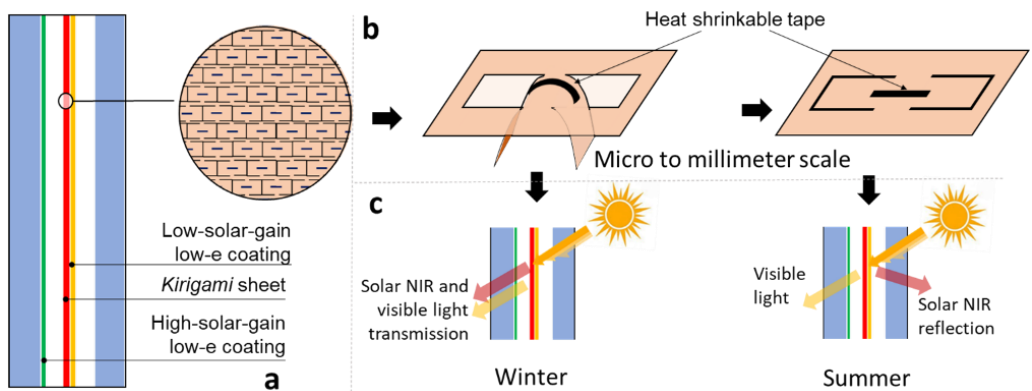

**Figure 2.** Schematics of the proposed window concept, structures, and mechanisms

To achieve such analyses, we used the Energy Management System (EMS) module in the EnergyPlus platform, which enables custom parametric energy simulations [19]. There are several components that make up an EMS program, including sensors, actuators, and calling points. In general, sensors extract information from the model, and they are leveraged to change a specific component or element of the model known as the actuator. Available sensors and actuators are provided in the RDD and EDD output files, respectively. There are many sensors relevant to simulating window technologies, such as indoor air temperature, relative humidity, and incident solar radiation, among others. Similarly, there are many building envelope actuators available. For this study, the actuator "Construction State" was utilized. The EMS program consists of EnergyPlus Runtime Language (erl) statements. Erl is a simplified programming language that supports IF statement structures and WHILE loops. The calling point determines when the EMS program is called with respect to the simulation. Further details on the EMS program developed for this study are provided in the next section.

**Table 1.** Double-pane window constructions used in a parametric simulation

| Case | Coating placement | SHGC | VT | U-Value (W/m$^2$-K) |
|---|---|---|---|---|
| Construction state 1 | Central layer between the panes | 0.19 | 0.30 | 1.56 |
| Construction state 2 | External surface of inner pane | 0.68 | 0.64 | 1.56 |
| Baseline_ Atlanta | - | 0.25 | 0.28 | 2.38 |
| Baseline_ Seattle | - | 0.36 | 0.40 | 2.04 |
| Baseline_ Denver | - | 0.38 | 0.42 | 2.04 |

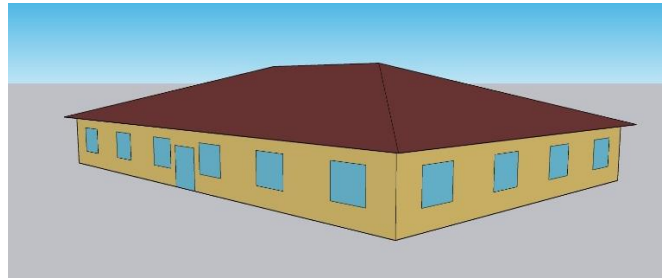

**Figure 3.** DOE prototypical building model for a small office building

This study used The Department of Energy (DOE) ASHRAE 90.1 2019 prototypical building model for a small office building (**Fig. 3**). The window-to-wall ratio ranges from 20-24% over the four facades. The parametric simulation was conducted in ASHRAE climate zones 3, 4, and 5, in Atlanta, GA, Seattle, WA, and Denver, CO, respectively. Based on the two selected low-e coatings and the opening ratio changes (0-0.8), we modeled the glazing system via the Lawrence Berkeley National Lab (LBNL) WINDOW software and obtained the spectral characteristics of the window constructions (see **Table 1**). As expected, the U-value remained constant, but the SHGC and VT increased due to the presence of the high SHGC low-e coating. These construction states were then used to simulate the proposed switchable between-glass built-in system through EnergyPlus EMS.

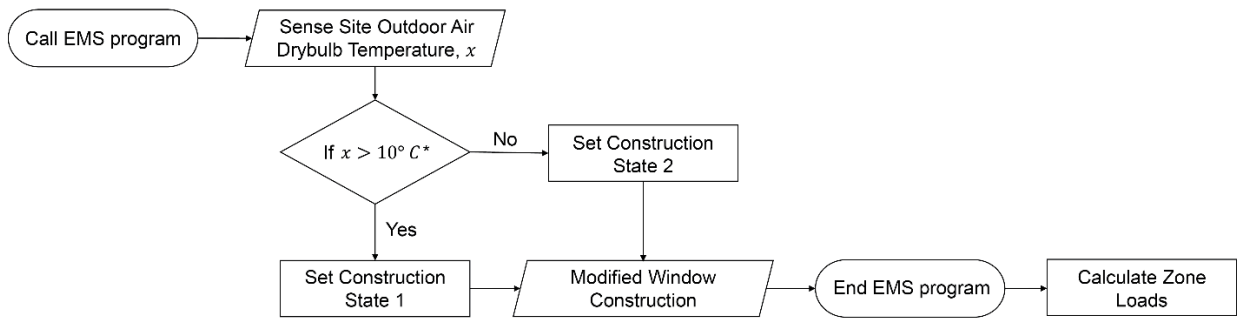

**Figure 4.** EMS workflow at each timestep during annual energy simulation

This study utilized the "Site Outdoor Air Drybulb Temperature" sensor and the "Construction State" actuator to modulate between two window constructions (**Table 1**). The program calling point was set to "Begin Timestep Before Predictor," which occurred near the beginning of each timestep but before zone loads were calculated. The simulations were conducted on an hourly resolution. As shown in **Fig. 4**, the EMS program determined if the temperature threshold was met and employed the corresponding construction state to all windows in the model. The parametric simulations were compared to the baseline simulations which used the prototypical window constructions provided in **Table 1**. It is worth mentioning that the final energy-saving performance can be influenced by the threshold temperature settings in the EMS simulation procedure, and there is a tradeoff between the high SHGC-driven heating energy savings and the low SHGC-driven cooling energy savings. In other words, an optimal temperature value may exist. However, the optimization study is not included in the scope of this work and will be explored in our future study. Rather, in this work, we explored several temperature values (ranging from 10 °C to 24 °C) as the thresholds for the EMS logic and also briefly evaluated their impacts on energy savings.

## 4. Results

### 4.1 Temperature threshold sensitivity analysis

A sensitivity analysis was conducted to select the threshold temperature for each climate zone (**Table 2**). For example, in climate zone 3, as the temperature threshold increased, the heating savings increased; however, the cooling savings significantly decreased, and the net energy savings became negative. In all three climate zones, 10 °C yielded the highest total energy savings.

**Table 2.** Temperature threshold sensitivity analysis

| Temperature (°C) | | 10 | 15 | 20 | 24 |
|---|---|---|---|---|---|
| Climate zone 3 | Heating savings % | 38.53 | 44.63 | 48.08 | 49.56 |
| | Cooling savings % | -0.25 | -3.64 | -8.59 | -13.69 |
| | Total savings % | 7.24 | 5.69 | 2.36 | -1.47 |
| Climate zone 4 | Heating savings % | 24.16 | 32.29 | 37.45 | 39.07 |
| | Cooling savings % | 14.52 | -1.43 | -25.30 | -41.30 |
| | Total savings % | 19.35 | 15.48 | 6.18 | -0.98 |
| Climate zone 5 | Heating savings % | 27.19 | 37.11 | 40.31 | 40.97 |
| | Cooling savings % | 10.88 | 3.46 | -7.10 | -23.08 |
| | Total savings % | 16.72 | 15.50 | 9.86 | 15.30 |

Although the optimization study was not included in the scope of this work, the above preliminary analysis of the temperature threshold effects demonstrates a tradeoff between the SHGC-drive heating energy savings and the cooling energy savings. Some parameters may have great impacts on the optimal temperature threshold value selection, including climatic conditions (e.g., ambient air temperature and solar irradiance), the building system's thermostat setpoints, baseline building envelope characteristics, and indoor heat gains. This would be explored in our future study.

### 4.2 Annual energy use analysis

Based on the confirmed temperature threshold values above, annual energy simulations were conducted in the three representative cities for both the baseline model (i.e., DOE prototypical building for a small office building) and dynamic model (i.e., DOE prototypical building for a small office building with dynamic NIR response). The annual heating and cooling loads were compared individually (**Fig. 5**), as well as the total energy consumption (**Table 3**).

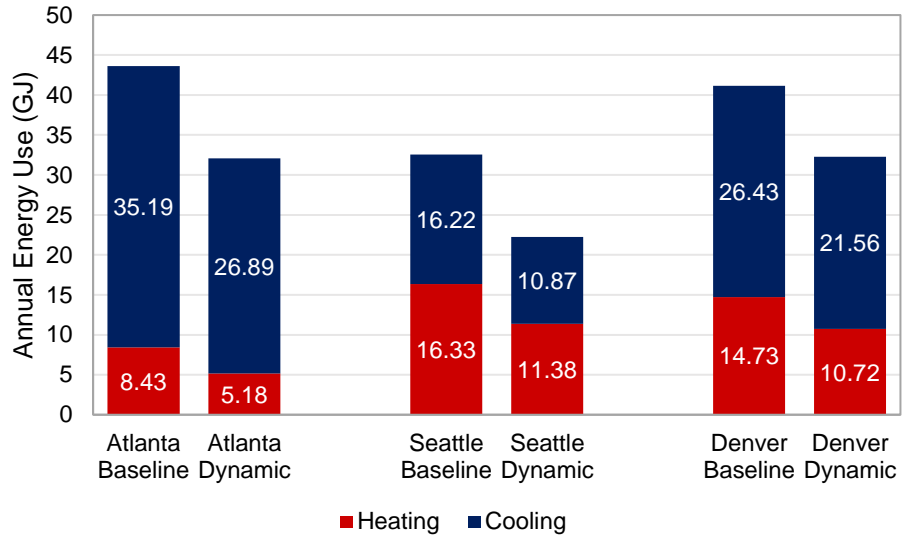

**Figure 5.** Annual heating and cooling load comparison

**Table 3.** Annual heating and cooling use and relative reduction due to changeable NIR response

| Climate zone | City | Baseline model annual energy use (GJ) | Dynamic model annual energy use (GJ) | Percentage % decrease (relative to baseline model) |
| --- | --- | --- | --- | --- |
| 3 | Atlanta, GA | 43.62 | 40.46 | -7.24 |
| 4 | Seattle, WA | 32.55 | 26.25 | -19.35 |
| 5 | Denver, CO | 41.16 | 34.28 | -16.72 |

Across the three climate zones, the dynamic model achieved ~7-19% annual energy savings. The dynamic model reduced the annual energy use in climate zones 3, 4, and 5 by 3.16 GJ, 6.30 GJ, and 6.88 GJ, respectively, which equated to 7.24%, 19.35%, and 16.72% deduction, respectively. Particularly the climate zone 4 model decreased the overall load by 19% and decreased the individual heating and cooling loads by 24% and 14%, respectively, compared to the baseline model. Meanwhile, the climate zone 5 model decreased the heating and cooling loads by 27% and 11%, respectively. Although the cooling load increased for the climate zone 3 model, heating energy savings were achieved, therefore the eventuate energy savings were still high. The results demonstrate a clear benefit of implementing this technology in mixed climates.

## 5. Conclusion

The proposed simulation method enables a comprehensive evaluation of the whole building energy effects of dynamic NIR modulations. A switchable between-glass built-in system was used to demonstrate the EMS-based framework. Using the proposed simulation method, the dynamic NIR-modulation technology achieved approximately 7% annual energy savings in climate zone 3, 19% in climate zone 4, and 17% in climate zone 5. The results provide insight into the effects of dynamically switchable NIR window technologies on heating and cooling loads among multiple climates. Additionally, refinement of the control strategy may yield greater energy savings using this method.

**Acknowledgements**

This research was funded by the National Science Foundation (2001207): CAREER: Understanding the Thermal and Optical Behaviors of the Near Infrared (NIR) - Selective Dynamic Glazing Structures and the USDA Natural Resources Conservation Service (NR203A750008G006): Spectrally Selective Solar Films for Operational Energy Savings of Urban Greenhouses.